\documentclass[12pt]{article}
\newcommand{\ra}{\ensuremath{\rightarrow}}

\newcommand{\ms}{\medskip}

\newcommand{\gke}{\ensuremath{\epsilon}}

\newcommand{\tcb}{\textcolor{blue}}

\usepackage{graphicx}
\graphicspath{ {./images/} }
\usepackage{sgamevar}

\usepackage{color}

\usepackage{latexsym}
\usepackage{amssymb}

\usepackage{latexsym}
\usepackage{amssymb}

   \newcounter{thm}

   \newcounter{def}

\newcounter{exm}

\newcounter{cor}

\begin{document}

 %\begin{frame}
\begin{center}
{\bf\Large A Model for the Capability Approach}\medskip

%\centerline{November 15, 2024}\medskip

\centerline{International Conference on Game Theory}

\centerline{Stony Brook, July 2025}
~\\
%\centerline{\bf Conference on Indian academic issues and IKS}\medskip

{\bf\large Rohit Parikh}

{City University of New York}

\end{center}

%~\medskip

%\end{frame} %\begin{frame}
\begin{abstract} {\small  ``The capability approach is a theoretical framework that entails two
normative claims: first, the claim that the freedom to achieve well-being is
of primary moral importance and, second, that well-being should be
understood in terms of people’s capabilities and functionings."  ~~Ingrid Robeyns and Morten Fibieger Byskov,~~{\em Stanford Encyclopedia of Philosophy}}\ms

  ``whether you should sleep on your back or on
your belly is a matter in which the society should permit you absolute freedom, 
even if a majority of the community is nosey enough to feel that you
must sleep on your back." ~~Sen 1970 \end{abstract}

%\end{frame} %\begin{frame}
\section{Introduction}

The capability approach was jointly founded by the Nobelist Amartya Sen, and Martha Nussbaum.  The motivation behind the capability approach is a dissatisfaction with the idea that a person's well being depends {\em solely}\, on what they {\em have}.  Rather, what Sen and Nussbaum  are suggesting is that what people {\em can do}\, with what they have is primary and what they have is only a tool to get what they want.  An example that Sen gives is of two people who have the same income but one of them is handicapped.   The one who is handicapped needs a wheelchair and the other one doesn't,  Then having the same amount of money is not an adequate comparison because a person who needs a wheelchair needs help and more resources in order to acquire the same well being as the other one.

%\end{frame} %\begin{frame}

What we propose to do in this paper is to point to two issues related to the capability approach.   

A)  One is that when one person's capability is increased then that fact may have negative consequences for someone else.  For example if I teach someone to pick a lock then he will acquire capabilities which he did not have earlier.   But on the other hand householders will then come under danger of theft or violence.  So one person's increase in capability may have negative consequences for another person. And this of course is the reason why many people object to the open carry law for weapons, which increases the capabilities of some people but puts some others in danger.

%\end{frame} %\begin{frame}

In 1893 in Southern Africa, British colonial police slaughtered 1,500 Ndebele warriors, losing only four of their own men in the process. This astronomical, almost unfathomable victory was earned not through superior strength, courage, or strategic skill, but because the British were armed with five machine guns and the Ndebele were not.  The invention and development of the machine gun by engineers such as Richard Gatling, William Gardner, and Hiram Maxim proved vital in colonization.

The extra capabilities of the British had a disastrous consequence for the Ndebele warriors, and eventually for Africans generally.

For a more contemporary example,  Thomas Matthew Crooks who shot Trump used a drone to survey the area of the Republican convention.  The drone increased his capability but brought a danger to Trump and to Corey Comperatore who died shielding his family from the gunman.

%\end{frame} %\begin{frame}

Of course, Sen and Nussbaum are primarily interested in helping those who are poor and need resources.  And machine guns were no doubt far from their minds.  But once we intrroduce the notion of capability, we do need to ask, {\em what kind of capability?  And capability for what?}.  When I teach a man to fish, I do increase his capability in some way.   He will be able to feed himself and his family.  But surely the fish are going to have a different view of that increase in capability.

The literature on the capability approach does not seem to explicitly acknowledge the game theoretic issues but these can clearly arise when one person's action increasing his utility may have the side effect of another person's utility becoming decreased.  And can the other person retaliate in some way?  We will return to this issue later on.

%\end{frame} %\begin{frame}

B)~  However a second issue that we want to address here is the following: namely what is {\em the quantitative measure}\, that goes with the capability approach?   For example let's suppose that a town has enough resources to build a swimming pool or alternately to have a bus line to a nearby city.  Now the swimming pool will obviously increase the capability of some people who want to be able to swim but cannot afford a pool at home.  The bus line will increase the capability of some other people who will then be able to travel to the nearby city to work there.  Which should the town do?  Obviously the town needs some measure that says that  the swimming pool will cause so much increase in capability and the bus line will result in some other increase in capability.  Which is greater?  And which sub-population will benefit?  The increases may not be for the same sub-population and we need some measure to decide whether to do the one or the other.  

%\end{frame} %\begin{frame}
\section{The Model}

In this model, there is a finite set $\cal A$ of agents, a set $\cal W$ of worlds.  Moreover each agent $i$ has a set of functions (capabilities) from $\cal W$ to $\cal W$.  The set of functions of agent $i$ will be denoted $C_{i,k}$ where $k$ ranges over the capabilities of agent $i$.  If an agent $i$ has capability $f$ then she can use $f$ to move (the world) from $w$ to $f(w)$.    Moreover for each world $w \in {\cal W}$ and each individual $i$ there is a value $v(i,w)$ which is the value of agent $i$ in world $w$.  If the agent $i$ is in the world $w$ and uses capability $f$ her value will change from $v(i,w)$ to $v(i,f(w))$.  If  $v(i,f(w))$ is larger than  $v(i,w)$ then she will carry out the action $f$.

%\end{frame} %\begin{frame}

   If Aditi is in a world $w$ where there is no ice cream, $f$ is the procedure of walking to an ice cream shop, then her value, by applying $f$, will increase from $v(a,w)$ to $v(a,f(w))$.  She will be happier and enjoy the ice cream.  And note that she needed not only the money to buy the ice cream but also the existence of the ice cream shop.  But the value of another agent $m$ who is Aditi's mother may also change as a result of Aditi applying $f$.  The mother will worry, ``where is my daughter?"   So $m$'s value may decrease.  We note that when Aditi changes \tcb{the world} she changes it both for herself and for her mother and even for the ice cream shop which acquires a tiny bit of revenue.

%\end{frame} %\begin{frame}

To avoid this problem we may require an {\em independence condition}\, that when an agent $i$ applies a capability $f$, only $i$'s value can change and no one else's value will change.  Aditi can walk to the ice cream shop without her mother worrying.  Thus if $f$ is a capacity of $i$ and $j$ is another agent, distinct from $i$ and $w$ is a world, then $v(j,w) = v(j,f(w)$.   We discuss a possible formalization of the independence condition in a later section.

%\end{frame} %\begin{frame}
Of course the real world is not like this and it could be that $v(j,f(w)) < v(j,w)$ even when $v(i,f(w))$ exceeds $v(i,w)$ .  But if some $f$ sharply increases the value for $i$ and the decrease in the value for $j$ is mild then we may consider $f$ to be all to the good.  If we take \$100 from a rich woman, and give it to a poor man, the rich woman's value (utility) will deccrease but only a little and the value of the poor man will increase sharply.  On the whole, \tcb{society at large} will be better off.  But we do need to be aware of the interdependence of value increase via procedures.

%\end{frame} %\begin{frame}

Now suppose that agent $i$ has capability $f$ which takes her from world $w$ to world $x$ and another capability $g$ which takes her from world $x$ to world $z$ then the composite capability $f \circ g$ will take her from $w$ to $z$.   After Aditi gets to the ice cream shop she will then have to perform other actions like ordering ice cream, paying for it, and eating it.  Thus typically agent $i$'s actual capabilities will be the closure of her basic capabilities under composition.  Let $C_i$ be this closure.  

%\end{frame} %\begin{frame}

Then agent i's capability net value in state $w$ is $V(i,w) = max(v(i,f(w)): f \in C_i)$.  If i's capabiliity set is increased to a larger set $D_i$ then her capability value $V$ will increase.  This is important to know. Do note the difference that $v(i,w)$ is agent i's {\em local value}\, at world $w$ whereas $V(i,w)$ is the maximum achievable by applying her capacity functions.  This distinction is important for Sen who asks us to focus on $V$ rather than on $v$.   Suppose that Shiva and Jack are two men with the same income and are at home drinking coffee.  Then we can suppose that $v(s,w) = v(j,w)$.  But suppose now that Shiva is handicapped and Jack is not.  And they both want to see the movie {\em Bajirao Mastani}\, at a theatre which is a mile away.  Then Jack's value $V(j,w)$ will be greater than Shiva's value $V(s,w)$,.  For Shiva can either not attend the movie, or can only do it at a substantial cost.

%\end{frame} %\begin{frame}

And this is where society comes in.  Suppose that the ice cream shop is too far for Aditi to walk to but the city has kindly provided a bus line which takes her from her home to the ice cream shop.  Then Aditi can perform the procedure $f$ which means walking to the bus stop, followed by procedure $b$, provided by society, i.e. procedure $f\circ b$ then she can enjoy her ice cream, assuming that the bus fare is small compared to the cost of the ice cream itself.  Let us denote her original capabiliity set as $C_i$ and the capability set after the addition of the bus line as $C_i + b$.  Then the net values are $V(i,w,C_i)$ and $V(i,w,C_i +b)$.  The second one is greater.

%\end{frame} %\begin{frame}

Thus the agent's capability value is actually dependent on three parameters.  Her identity $i$, the world $w$ where she is now, and her capability set $C_i$.  When society provides an additional procedure $b$ to $i$ which $i$ can use, then her capability set changes to the larger set $C_i + b$.  The gain to her from $b$ is 

$$G(i,w,b) = V(i,w,C_i+b) - V(i,w,C_i)$$

Now we note that the gain to an individual $i$ may be different from the gain (or loss) to another individual $j$.  Someone with a car will not be helped much by a new bus line.  A poor person without a car will not be helped by a new gas station.

%\end{frame} %\begin{frame}

When we are comparing two possible social procedures $p$ and $q$ we will need to consider {\em who}\, is helped by $p$ and who is helped by $q$.    Generally speaking, we will seek to help those $i$ for whom their current value is small rather than others $j$ whose current value is already large.   But such issues are for another paper.  

we now turn to the issue of how a  procedure which helps one person may harm another one.

%\end{frame} %\begin{frame}

 \section{A Game theoretic example}

This example is simple but illustrates the main point that one person's increase in capability can be harmful to another person.

In this example, Sona is considering selling her car to Ravi.  Sona bought the car for \$500 and is offering it for \$600.  The car is worth \$900 to Ravi who can use it to meet customers.  

Here Ravi is the row player and Sona is the column player.  Sona can offer the car or not offer.  Ravi can offer to buy or not offer.

\begin{center}
  \begin{game}{2}{2}
      \> sell   \> no sell \\
    buy \> (300,100)  \> (0,0) \\
  no buy \> (0,0) \> (0,0) ~   
 
  \end{game}\vspace{.2in}
\end{center}

It is in the interest of both that the sale takes place.  (buy, sell) is a Nash equilibrium.

%\end{frame} %\begin{frame}

But suppose now that Ravi acquires a gun and threatens Sona if she does not give the car for free.

\begin{center}
  \begin{game}{3}{2}
      \> give   \> no give \\
    buy \> (300,100)  \> (0,0) \\
  no buy \> (0,0) \> (0,0) \\   
  threaten Sona \> (900,-500) \> (0,-1,000)~
  \end{game}\vspace{.2in}
\end{center}

Now {\em threaten Sona}\, is the dominant strategy for Ravi.  The new game has increased Ravi's capability but makes things worse for Sona.  The new Nash equilirium (threaten, give) is much better for Ravi.  We {\em could}\, now bring in the legal system which threatens Ravi with prison and the Nash equiibrium will return to the original (buy, sell).  In order for that to happen, the expected value of threatening plus punishment by society must be less than 300 for Ravi.

%\end{frame} %\begin{frame}

{\bf\large A controversial example}

{\small Suppose the chairman of a department wants to make a proposal $P$.  the proposal is that in the future the common room of the department should only have tea and no coffee should be served.  He expects four faculty to be present at the next department meeting and he knows that only one of the other faculty supports the tea only proposal. But the other three are opposed to it, so if the proposal is put to a vote it will lose.  

However the chairman anticipating this problem comes to the meeting with three friends who he claims are members of the department.  The three friends all vote in favor of tea and so the proposal passes.  At this point one of the four original faculty wants to know,  ``are you quite sure that these three friends of yours are actually faculty?"

The chairman says ``yes they are members of the faculty.  You may not have met them earlier but they are indeed members."   Then the person who asked the question says ``may I see their ID?" and the chair says ``you know very well that asking for an ID is undemocratic and you should not ask for such a shameful thing "}

%\end{frame} %\begin{frame}

So let us suppose that the three friends of the chairman were not actually faculty but what is the harm if they are given some rights?   Their capability is increased and surely that is good?   But the harm is that by showing up at the meeting and voting without having the right to do so, they have {\em diluted}\, the votes of the regular faculty.  They helped to pass proposal P when in fact it should not have passed.

The point is that when I vote I am not merely expressing my own opinion.  I am also diluting the votes of the other people whose views may be different from mine.  So every vote has both a positive and a negative vote effect.  The positive effect is that it expresses my own views and the negative effect is that it dilutes the views of the others.

%\end{frame} %\begin{frame}

 It tends to be Republicans who are very concerned with the ID requirement and they are criticized with the argument that the Republicans are opposed to democracy.  And of course it goes without saying that a person who is a faculty member \tcb{is entitled to vote} and a person who is a citizen is entitled to vote in the general election.   What if someone who is actually a faculty member forgot her ID at home?  Should she be deprived of her right to vote? What if a person who is a citizen has not got the means to get an ID?  Should he be prevented from voting?    These issues are also relevant.

But at the same time we can see that the ID requirement is an important requirement to make sure not merely that those who are entitled to vote do vote but also to make sure that those who are not entitled to vote do not end up diluting the votes of those who are entitled. 

%\end{frame} %\begin{frame}

{\bf\large $\epsilon$ is not zero}

There are many contexts in which $\epsilon$ is used as a symbol for a quantity which is small but positive.  In social applications, how do we use it?  Do we use it as something which does not matter?  Or as something which matters but is much less than other {\em real} quantities?    We define the derivative of $f$ at $a$ by the formula 
{\large
$$f'(a)  = lim_{ \gke \ra 0}(f(a+\gke) - f(a))/\gke$$}

%\end{frame} %\begin{frame}

And here it does not matter which number is $\gke$.  It is small and going to zero.   But elections often involve small numbers repeated many times and so it does matter {\em how small}  and {\em how many}.  And in other contexts as well.
Suppose that a beer company puts up an advertisement in a subway station.  The effect on each person  who sees it will be small, but many people see it.    To justify their cost, they have to multiply one small number by another, large number.    They will advertise if the small effect on each passenger, multiplied by the large number of passengers, exceeds their cost.  A similar issue arises with the {\em Tragedy of the Commons} where a small gain to each herder results in a large loss to the entire community.

So let us consider the game theoretic situation when someone A, who is not entitled to vote, manages to vote anyway.

%\end{frame} %\begin{frame}

Clearly there is a positive payoff to A by voting, but it takes time to go and vote, one may have to stand in line.  The satisfaction must exceed these costs.  So let us say that $\gke$ is the net gain to A.  There is presumably a loss to society, for otherwise why not allow everyone, even children, to vote?   Let us assme that of the two alternatives, one is good $G$ and one is bad, $B$.  And $h$ is the harm to society if $B$ wins.  But that harm must be multiplied by the probability of A's vote \tcb{actually} causing $B$ to win.  For that to happen, the votes for $G$ and $B$ must be equal.  What are the chances of that?

%\end{frame} %\begin{frame}

Now in many states, red states and blue states, one party has such an overwhelming advantage that one vote cannot possibly make a difference.  But there are indeed purple states.  As NPR's Domenico Montanaro has put it, in 2020, {\em just 44,000 votes in Georgia, Arizona and Wisconsin separated Biden and Trump from a tie in the Electoral College.}\medskip

%\tcb{\large This is where my question is.}  How big is the number?

So what is the probability that the 44,000 votes would be evenly divided? That number is $(22000!\times 22000!)/44000!$.
A rough calculation shows that the number is less than $0.75^{(11,000)}$ since the ratio includes 11,000 factors ranging from 0.50 to 0.75.  Thus the harm done to society by one person voting illegally is actually very very small.   The situation of course would be different if several people were to vote illegally.  George W.  Bush won Florida's electoral votes by a margin of only 537 votes out of almost six million cast.

%\end{frame} %\begin{frame}
\section{Other considerations}

The next two subsections give only a hint of what developments can happen next.

 {\em Formalizing independence}  When we considered Aditi walking to an ice cream shop to buy ice cream, we also considered the worry caused to her mother who wonders where Aditi is.  But suppose we focus \tcb{only} on the pleasure to Aditi.  We could consider a model where each agent $i$ has her own world $W_i$ and the large world is ${\cal W} = \Pi\, W_i, i\in A$.  A function applied by agent $i$ may alter $W_i$ but no other world.  I assume that such an idea is behind what Sen suggested that each man is entitled to paint his bathroom any color that he likes.  His bathroom is \tcb{his} domain and no one else has anything to say about it.   Actually Sen does not assume independence in his example of the book {\em Lady Chatterly's Lover}, because \tcb{at most one} of Prude and Lewd can read the book.

%\end{frame} %\begin{frame}

\section{Sen's original example}

Sen's original example used a simple society with only two people and only one social issue to consider. The two members of society are named "Lewd" and "Prude". In this society there is a copy of a {\em Lady Chatterley's Lover} and it must be given either to Lewd to read, to Prude to read, or disposed of - unread.

 Suppose that Lewd enjoys this sort of reading and would prefer to read it rather than have it disposed of. However, he would get even more enjoyment out of Prude being forced to read it.

Prude thinks that the book is indecent and that it should be disposed of, unread. However, if someone must read it, Prude would prefer to read it rather than Lewd since Prude thinks it would be even worse for someone to read and enjoy the book rather than, as Prude herself would do, read it in disgust.

%\end{frame} %\begin{frame}

Given these preferences of the two individuals in the society, a social planner must decide what to do. Should the planner force Lewd to read the book, force Prude to read the book or let it go unread? More particularly, the social planner must rank all three possible outcomes in terms of their social desirability. 

The social planner decides that he should be committed to individual rights, and each individual should get to choose whether they, themself will read the book. Lewd should get to decide that the outcome "Lewd reads" will be ranked higher than "No one reads", and similarly Prude should get to decide that the outcome "Prude reads" will be ranked lower than "No one reads".

%\end{frame} %\begin{frame}

Following this strategy, the social planner declares that the outcome "Lewd reads" will be ranked higher than "No one reads" (because of Lewd's preferences) and that "No one reads" will be ranked higher than "Prude reads" (because of Prude's preferences). 

Consistency then requires that "Lewd reads" be ranked higher than "Prude reads", and so the social planner gives the book to Lewd to read.

But, this outcome is regarded as worse than "Prude reads" by both Prude and Lewd, and the chosen outcome is therefore {\em Pareto inferior}  to another available outcome—the one where Prude is forced to read the book.  And this violates Pareto optimality.

\subsection{Indpendence}   Suppose (for simplicity) that there are three agents \{1,2,3\}, each has her own world $W_i$ and the global world $\cal W$ is simply the cross product $W_1\times W_2 \times W_3$.  Moreover the utility of agent i depends entirel in the i'th coordinate.  Thus for instance the utility of agent 2 is the same for worlds $(a,b,c)$ and $(a,b',c)$.  In that case Sen's condition  ``whether you should sleep on your back or on your belly is a matter in which the society should permit you absolute freedom" can be fulfilled.  Whether you sleep on your back or on your belly affects no one else's utility and they should not care.   

But one can imagine that this condition might not hold.  Your wife, who sleeps besife you, is aware that you snore when you sleep on your back but not when you sleep on your belly.  Then her  utility is going to be affected by your choice.  For a more serious example, suppose you have been injured and are in a hospital.  Your recovery can proceed only if you sleep on your belly.  In that case the nurse taking care of you is not going to be indifferent to how you sleep.  If a butterfly flapping its wings in Brazil can cause a snowstorm in Boston, then we may well care very much what the butterfly does with its wings.

%\end{frame} %\begin{frame}

It is clear that the notion of independence is crucial in how we live our lives.  Most of the time we try to amend our personal world $W_i$ but the inevitable interdependence can frustrate our plans.    Our plan to drive to New Jersey can be frustrated if the bridge is closed.  And the lion will go hungry if the deer fulfills her wish to stay alive.

%\end{frame} %\begin{frame}

\section{Regulations}  It may well happen that some of our inherent capabilities can be blocked by a social convention or by a law. If $C_i$ is some naturally arising set of capabilities for an agent $i$ but there are laws which restrict some of the functions $f$ available to $i$, then $C_i$ may reduce to a smaller $C^-_i$ and decrease the value function $V$ for agent $i$.  But $i$ may benefit if another person's capabilities are reduced by his not being able to carry his gun in $i$'s children's school room.  There is clearly a trade off here where a loss to one means a gain to another.   This investigation also requires a further technical development but we do not here have the space to address it.

%\end{frame} %\begin{frame}

{\bf\large Conclusion}

We have proposed a preliminary model to think technically about capabilities.  We have also brought into prominence certain issues which we have not seen addressed elsewhere.  An important one is that one person's gain may be another person's loss.  So an increase in capability may not always be an unmixed blessing.  The other is that when society provides some means of increases capability then some sub-populations may benefit more than others.  When society allows you to deduct the value of your yacht from your income tax, then the poor or the middle class will not benefit.   

%\end{frame} %\begin{frame}

The development in this version is very preliminary and we hope to expand on these themes. 
Some of the references like those by Sen and Nussbaum are basic to this paper.  Others refer to the tools which will come in during further development.\medskip

But an important point, once we take the capability approach seriously, is to realize that the infrastructure of society is part of the ``wealth" of its members.  New York city has a large public transportation system and travelling by subway is cheaper and more convenient than driving a car.  I would suggest that a well run subway system is part of the well being of the residents, and a reasonable amount should be added to the per capita income of New York city residents.   Of course we are aware that the subway system is a public good.  But we need also to assign a {\em number}\, which is added to the per capita income.

%{\bf Acknolwedgement:}  thanks to Paul Pedersen for assistance with formatting.

%\end{frame} %\begin{frame}

\section{References}

\begin{itemize}

{\footnotesize

\item {\em The Cambridge Handbook of the Capability Approach},
Edited by Enrica Chiappero-Martinetti, Siddiqur Osmani, Mozaffar Qizilbash, Cambridge University Press, 2021.

%\item Cormen, Thomas H., et al. {\em Introduction to algorithms}. MIT press, 2022.

\item  van Hees, Martin. "Analyzing Capabilities." The Cambridge Handbook of the Capability Approach (2020).

\item Nussbaum, Martha C. {\em Creating capabilities: The human development approach}. Harvard University Press, 2011

\item Parikh, Rohit. "Social software." {\em Synthese}\, 132 (2002): 187-211.

%\item Pratt, Vaughan R. "Semantical considerations on Floyd-Hoare logic." 17th Annual Symposium on %Foundations of Computer Science (sfcs 1976). IEEE, 1976.

\item Robeyns, Ingrid. "The capability approach: a theoretical survey."  {\em Journal of human development} 6.1 (2005): 93-117.

\item Ingrid Robeyns and Morten Fibieger Byskov, The Capability Approach, {\em Stanford Encyclopedia of Philosophy}

\item Sen, Amartya (1970). "The Impossibility of a Paretian Liberal" {\em Journal of Political Economy}. 78 (1): 152–157.

\item Amartya K. Sen, ``Rational Fools: A Critique of the Behavioral Foundations of Economic Theory"
{\em Philosophy \& Public Affairs}, Vol. 6, No. 4 (Summer, 1977), pp. 317-344 (28 pages)

\item  Sen, Amartya. "Human rights and capabilities." {\em Journal of human development} 6.2 (2005): 151-166.

}
\end{itemize}
\end{document}